\documentclass[
aps,%
prd,
12pt,%
final,%
notitlepage,%
oneside,
onecolumn,
nobibnotes,%
nofootinbib,%
superscriptaddress,%
noshowpacs,%
centertags]%
{revtex4-2}

\usepackage{tikz}
\usepackage[mathscr]{eucal}
\usepackage{hyperref}
\usepackage{mathtools}
\usepackage{amsmath}
\usepackage{amssymb}
\usepackage{cancel}
\DeclareMathOperator{\sign}{sign}

\begin{document}
	
	\title{Supersymmetric proof to count bound state nodes}
	
	\begin{abstract}
		A normalizable static supersymmetric bound ground state annihilated by the super-generators has got zero number of internal nodes in the framework of one-dimensional supersymmetric quantum mechanics. The super-generator transformations between excited super-partner bound states as combined with the standard technique of Wronskian provides an elegant and self-sufficient way to derive the equality of internal nodes amount to the number of consequent excitation.
	\end{abstract}

	\author{\firstname{Asya}~\surname{Aynbund}}
	\email{aynbund.asya@phystech.edu}
	\affiliation{Landau Phystech-School, Moscow Institute of Physics and Technology,
		Russia, 141701, Moscow Region, Dolgoprudny, Institutsky 9}
	
	\author{V.V.Kiselev}
	\email{kiselev.vv@phystech.edu; Valery.Kiselev@ihep.ru}
	
	\affiliation{Landau Phystech-School, Moscow Institute of Physics and Technology,
		Russia, 141701, Moscow Region, Dolgoprudny, Institutsky 9} 
	\affiliation{State
		Research Center of the Russian Federation  ``Institute for High Energy
		Physics'' of National Research Centre  ``Kurchatov Institute'',
		Russia, 142281, Moscow Region, Protvino, Nauki 1}
	
	\maketitle

\section{Introduction \label{sec:Int}}

Counting an amount of internal nodes in wave functions of bound states is known as an oscillation theorem. The theorem relates a sequence number of excited state in the order of energy enlarge to number of its nodes. An ordinary proof, say, in classical book by F.\,Berezin and M.\,Shubin~\cite{BerezinShubin} is a set of theorems, lemmas, consequences and prepositions in Sturm--Liouville theory of differential equations. Such the way is rigorous and it looks like just straightforwardly mathematical, while physical explanations merely give some qualitative arguments but not enough to be provements (see, for instance,~\cite{nodes}). In occasion there is a method based on coherent quantum structure, that provides an elegant and physically clear way to derive the oscillation theorem by means of supersymmetry in one dimensional quantum mechanics. This method is built on a construction introduced by M.\,Crum \cite{Crum}  and incorporated to the supersymmetry by E.\,Witten~\cite{Witten} as reviewed in~\cite{SQM,UFN,Phys.Rep}.  

In Section~\ref{sec:MachSQM} basic notions and constructions are described to give the most important properties of supersymmetric Hamiltonian. Section~\ref{sec:GrSt_inDS} deals with a ground state, and we derive that the ground state has got no internal nodes. Section~\ref{sec:CountNodes} is devoted to a step to the level next to the ground state to prove that the number of nodes increases to the unit exactly. Two Appendices \ref{sec:AppA} and \ref{sec:AppB} are devoted to some ordinary details for a completeness of presentation  as concerns for proofs of both a bound state wave function to be real and an amount of nodes to increase with growing an excitation number. 
In Conclusion we summarize the results. 

\section{Machinery of Supersymmetric Quantum Mechanics \label{sec:MachSQM}}

The theory under study is one-dimensional stationary Schr\"odinger equation. Suppose that  $\Psi_0$ is a ground state wave function normalised in a Hilbert space of quantum states for a Hamiltonian $\hat H_1$. As it is well known, function  $\Psi_0$  can be set to be real since it has got nodes at borders (see a standard derivation in Appendix \ref{sec:AppA}). Let the Hamiltonian have got several excitations of the ground state, and the excited levels are counted in the order of energy enlargement as $\Psi_k^{(1)}$. Following M.\,Crum~\cite{Crum} and M.\,Darboux~\cite{Darboux}, one constructs a partner Hamiltonian $\hat H_2$, that possesses a set of bound states wave functions $\Psi_k^{(2)}$. 

Define two linear differential operators $A^-$ and $A^+$, which are hermitian conjugated each to other:
$$
A^-=-\frac{d}{dx}+\frac{\Psi'_0}{\Psi_0},\qquad 
A^+=\frac{d}{dx}+\frac{\Psi'_0}{\Psi_0},
$$
where the symbol of prime denotes the derivative with respect to $x$.

Introduce operators of super-generators $Q$ and $\bar{Q}\equiv Q^\dagger$ by E.Witten~\cite{Witten}
$$
Q=\frac{\hbar}{\sqrt{2m}}
\begin{pmatrix}
	0 & 0 \\
	A^- & 0 \\
\end{pmatrix} 
\text { and } 
\bar{Q}=\frac{\hbar}{\sqrt{2m}}
\begin{pmatrix}
	0 & A^+ \\
	0 & 0 \\
\end{pmatrix} 
$$
\text { acting on } 
$$
\begin{pmatrix}
	\Psi^{(1)} \\ \Psi^{(2)}
\end{pmatrix}
$$
where the column refers to wave functions of super-partner Hamiltonians $\hat H_1$ and $\hat H_2$, respectively.

A supersymmetric Hamiltonian $\mathcal{H}$ is constructed as
$$
\mathcal{H} =\left\{Q, \bar Q\right\}.
$$
Consider  properties of Hamiltonian $\mathcal H$ and supercharge operators. So, the super-generators are nilpotent, i. e. $Q^2=\bar Q^{2}=0$ and they commute with Hamiltonian
$$
[Q, \mathcal{H}]=0.
$$
Then the definition of super-partner Hamiltonians in terms of matrix $\mathcal H$ reads off

$$
\mathcal{H}=\frac{\hbar^2}{2m}\begin{pmatrix}
	A^+A^- & 0 \\
	0 & A^-A^+ \\
\end{pmatrix} =\begin{pmatrix}
	\hat H_1-E_0& 0 \\
	0 & \hat H_2-E_0 \\
\end{pmatrix}.
$$
Two Schr\"odinger equations under study are	
$$
\hat{H_1}\Psi^{(1)}_k=E_k\Psi^{(1)}_k,\qquad 
\hat{H_2}\Psi^{(2)}_k=E_k\Psi^{(2)}_k,
$$
where the eigenvalues of Hamiltonians coincide. Indeed, according to the connection between $\hat{H_1}$ and operators $A^-$ and $A^+$ the action of $A^+A^-$ to the wave function  $\Psi^{(1)}_k $ gives	
$$
\frac{\hbar^2}{2m}A^+A^-\Psi^{(1)}_k = (\hat{H_1} - E_0)\Psi^{(1)}_k = 
(E_k - E_0)\Psi^{(1)}_k .
$$
Acting by $A^-$ results in the eigen state equation
\begin{multline*}
\frac{\hbar^2}{2m}A^-A^+(A^-\Psi_k^{(1)}) = (E_k-E_0)(A^-\Psi_k^{(1)})\quad
\Rightarrow\quad\\ {}
(\hat H_2-E_0)(A^-\Psi_k^{(1)})=(E_k-E_0)(A^-\Psi_k^{(1)}),
\end{multline*}
determining the wave function $\Psi_k^{(2)} \stackrel{\text{def}}{=} A^-\Psi_k^{(1)}$ as the bound state solution for  Hamiltonian $\hat H_2$. Note that the superscripts of wave functions point to means the super-partner Hamiltonians, while the subscript denotes the number of excitation with the same energy $E_k$ for both Hamiltonians.

\section{Ground state in the discrete spectrum \label{sec:GrSt_inDS}}

The ground state wave function $\Psi_0$ satisfies the equation of ground state:
$$
A^- \Psi_0 = 0.
$$
Hence, any other function satisfying the same equation is a normalised solution for the ground state and it is proportional to $\Psi_0$, since the bound state energies are not degenerate. 

Let us show that $A^-|\Psi_0|=0$. Therefore, $$\Psi_0 = \mbox{const}\cdot|\Psi_0|,$$ and generically one can put the constant to be positive, while  $\Psi_0>0$ at any internal point except borders. It means that $\Psi_0$ is nodeless. 

Indeed, 
$$
\tilde{\Psi} \stackrel{\text{def}}{=} |\Psi_0| = \Psi_0 \sign (\Psi_0),
$$
while the action by $A^-$ straightforwardly gives
\begin{multline}
	A^-\tilde{\Psi} =\left(-\frac{d}{dx}+\frac{\Psi'_0}{\Psi_0}\right)\Psi_0 \sign (\Psi_0) = 
	\\ {} =-\frac{d}{dx}\Big(\Psi_0 \sign (\Psi_0)\Big) + \Psi'_0 \sign (\Psi_0) = \\			{}
	= \cancel{-\sign (\Psi_0) \Psi'_0} - \Psi_0 \frac{d}{dx}\Big(\sign (\Psi_0)\Big)+\cancel{\Psi'_0 \sign (\Psi_0)} =\\ {} =   - \Psi_0 \frac{d}{dx}\Big(\sign (\Psi_0)\Big) = \\ {}
	= -\sum_{j} \Psi_0(x_j)\,2\, \delta(x_j)\,\sign\hskip-2pt\Big(\Psi'_0(x_j)\Big) \equiv 0,\qquad 
\end{multline}
where $x_j$  are positions of possible nodes of $\Psi_0$, if exist, while $\Psi'_0(x_j)\neq 0$ because $$\Psi'(x_j)=\Psi(x_j)=0$$ would lead to $\Psi(x)\equiv 0$ as a solution of Schr\"odinger equation. 

Thus, the super-generators provide us with the short proof of statement that the ground state wave function $\Psi_0$ is nodeless, so we consider it to be positive anywhere except border points.

\section{Counting the nodes \label{sec:CountNodes}}

\begin{figure}[t!]
	\centering
	\includegraphics[width=0.45\textwidth]{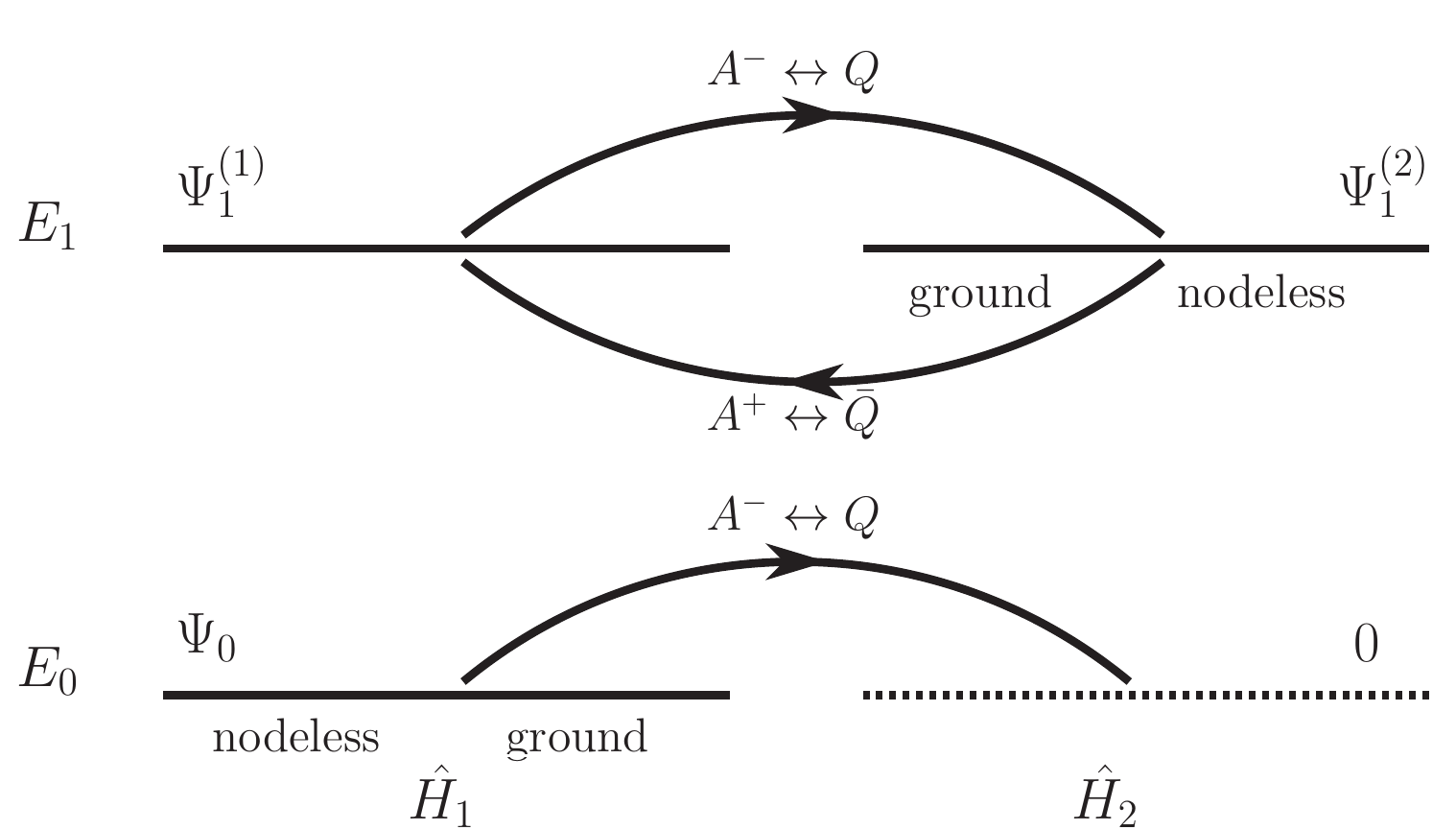}
	\caption{The ground state and the exited level of Hamiltonian $
		\hat H_1$ (left), the ground state of Hamiltonian $\hat H_2$ (right) 
		with energies $E_0$ and $E_1$, respectively.} 
	\label{fig:levels}
\end{figure}

As we plot in Fig.~\ref{fig:levels}, there are two sets of states for the super-partner 
Hamiltonians with energy $E_1$. We show that one can transform one column of states to another by making use the linear operator $A^-$ for moving from left to right (for example, if one acts on  $\Psi_0$, we will get $0$) and the operator $A^+$ for moving back. These operators conserve the energy level. \par The figure points that the ground states for both super-partner Hamiltonians are nodeless, while more energy means that the wave function has more nodes as well known fact. So, with this explicit scheme for the super-partner system we are going to prove that the first excitation of ground state for $\hat H_1$ has got exactly single internal node, while further excitations have got numbers of nodes equal to its excitation numbers. 

In previous section we proved so-called \textit{base case} of mathematical induction method. The next point is to prove the induction step. Acting on the first excited wave function of Hamiltonian $\hat H_2$ and by revealing the operator by definition we get
\begin{multline*}
	A^+\Psi_1^{(2)} =\Psi_1^{'(2)} + \frac{\Psi'_0}{\Psi_0}\Psi_1^{(2)}  = \frac{\left(\Psi_0\Psi_1^{(2)}\right)'}{\Psi_0} = \langle \text { or } \rangle
	= \\ {} =A^+(A^-\Psi_1^{(1)}) =  \frac{2m}{\hbar^2}(E_1-E_0)\Psi_1^{(1)}.
\end{multline*}
Then, multiplying by $\Psi_0$ one can see the derivative of Wronskian for $\Psi_0$ and $\Psi_1^{(1)}$:
\begin{multline*}
\left(\Psi_0\Psi_1^{(2)}\right)' =  \frac{2m}{\hbar^2}(E_1-E_0)\Psi_1^{(1)}\Psi_0 = - W'_{\Psi_0\Psi_1^{(1)}} \Rightarrow \\ {} - W_{\Psi_0\Psi_1^{(1)}}=\Psi_0 \Psi_1^{(2)}.
\end{multline*}
Therefore, we come to the following expression:
$$
{\Psi_0'{\Psi_1^{(1)}}-{\Psi_0}\Psi_1'^{(1)}=\Psi_0 {\Psi_1^{(2)}}.}
$$
Consider the first term on the left side. Under the condition of $E_1$ being greater than $E_0$ one concludes that an amount of nodes in  the excited level wave function $\Psi_1^{(1)}$ is strictly greater than the amount of nodes in the lower level wave function $\Psi_0$ (see Appendix \ref{sec:AppB}). Hence, there is at least one intrinsic node which we denote by $x_1$. Rewriting the previous equation at the node we get
$$
\Psi_0(x_1) \Psi_1^{(2)}(x_1)=-\Psi_0(x_1)\Psi_1'^{(1)}(x_1), 
$$ 
where all of $\Psi_0(x_1)$ and $\Psi_1^{(2)}(x_1)$  are positive by construction for the ground states (remember, that $\Psi_1^{(2)}$ is the ground state for Hamiltonian $\hat H_2$). 

\begin{figure}[t!]
	\centering
	\includegraphics[width=0.47\textwidth]{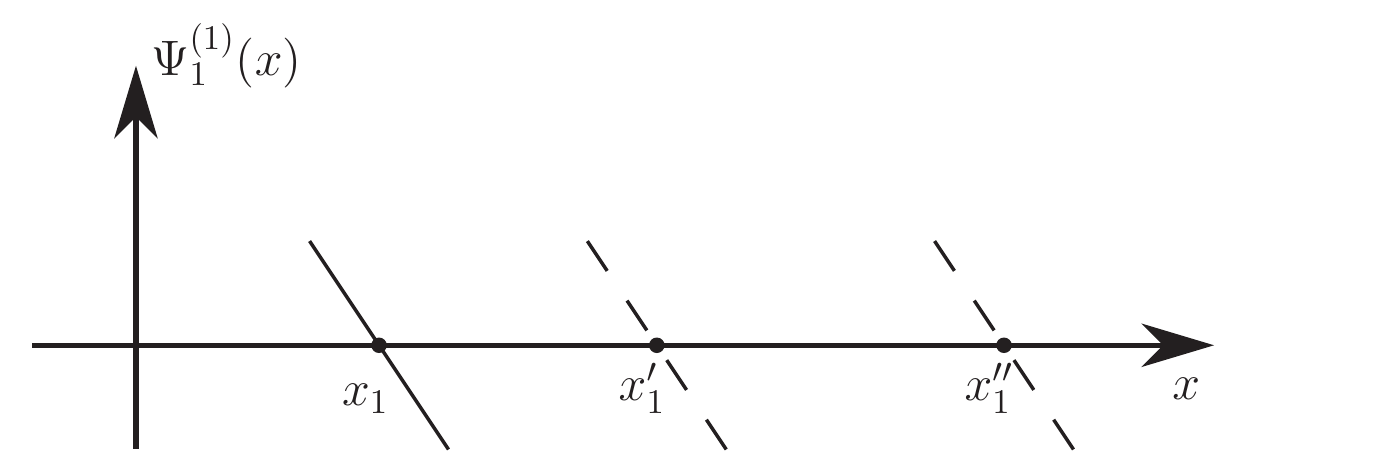}
	\caption{Node of the exited state and the direction of the derivative in it.} 
	\label{fig:x_1}
\end{figure}

Therefore, $\Psi_1'^{(1)}(x_1)<0$, as it is depicted in Fig.~\ref{fig:x_1}, for any node $x_1$ if more than one. It means that in all of nodes the derivative has to be negative. Since the wave function is continuous, we have to conclude that there is a single change of  $\Psi_1^{(1)}$ sign only, and the wave function of the first excitation $\Psi_1^{(1)}$ has a single internal node, exactly.

Further steps of mathematical induction are quite transparent.

\section{Conclusion}

In the present paper we have presented the new original modern proof for counting the internal nodes of bound states wave functions instead of Sturm--Liouville theory by making use the elegant mathematical formalism of Supersymmetric Quantum Mechanics. It could be incorporated into textbooks on modern quantum mechanics in order to provide some acquaintance with up-to-date supersymmetric concepts and mathematical methods essential in nowadays science. 

In practice the counting of nodes is actual in researches dealing with spectra of exotic tetraquark and pentaquark states in QCD.
	
	 \clearpage
	\appendix
	\section{Single wave function of bound state \label{sec:AppA}}
A wave function of bound state in the one-dimensional quantum mechanics has got nodes at border points $x_b$: $\Psi(x_b)=0$, that ensures the probability current $j$ to be equal to zero at the border points, since
$$
j=-\frac{\mathrm{i}\hbar}{2m}\Big(\Psi^*(x)\Psi^\prime(x)-
\big(\Psi^\prime(x)\big)^*\Psi(x)\Big)\Big|_{x=x_b}\equiv 0.
$$	
This generic condition makes definite conclusions for a Wronskian
$$
W_{\Psi_n\Psi_k} \stackrel{\text{def}}{=} 
\det\begin{pmatrix}
	\Psi_n & \Psi_k\\ 
	\Psi'_n & \Psi_k'
\end{pmatrix} =\Psi_n\Psi_k' - \Psi_k\Psi'_n.
$$
defined by two different bound solutions of Schr\"odinger equation ${\hat H}\Psi_k = E_k\Psi_k$ with the same Hamiltonian $\hat H$. So,

\begin{equation}
	\label{a1-1}
	W_{\Psi_n\Psi_k}\Big|_{x=x_b}\equiv 0.
\end{equation}
Moreover, the Wronskian derivative equals 
\begin{multline}\displaystyle 
	W'_{\Psi_n\Psi_k} =
	\frac{\mathrm{d}}{\mathrm{d}x}\det\begin{pmatrix}
		\Psi_n & \Psi_k\\ 
		\Psi'_n & \Psi_k'
	\end{pmatrix} 
	= \left(\Psi_n\Psi_k' - \Psi_k\Psi'_n\right)' =\\ {} \displaystyle =
	\cancel{\Psi'_n\Psi_k'} - \cancel{\Psi_k'\Psi'_n} + \Psi_n\Psi_k'' - \Psi_k\Psi''_n  = \\ {} =\Psi_n\Psi_k'' - \Psi_k\Psi''_n. \label{a1}
\end{multline}
Under  Schr\"odinger equations
$$
\begin{array}{l}\displaystyle 
	\Psi''_n =\frac{2m}{\hbar^2}\big(V- E_n\big)\Psi_n,\\[2mm]
	\displaystyle 
	\Psi''_k =\frac{2m}{\hbar^2}\big(V- E_k\big)\Psi_k,
\end{array}
$$
Eq. (\ref{a1}) gets the form  
$$
W'_{\Psi_n\Psi_k} = \frac{2m}{\hbar^2}
(E_n - E_k)\Psi_n \Psi_k.
$$
If $E_n=E_k$ we put $\Psi_k=\tilde\Psi_n$, then 		
$$
W'_{\Psi_n\tilde \Psi_n} \equiv 0.
$$
Therefore,
$$
W_{\Psi_n\tilde \Psi_n}=\mbox{const.}
$$	
The constant is determined  by the condition at the border point (\ref{a1-1}), that results in 
\begin{equation}
	\label{a1-2}
	W_{\Psi_n\tilde \Psi_n}\equiv 0\quad \Rightarrow\quad
	\Psi_n\tilde \Psi_n' -\tilde  \Psi_n\Psi'_n=0.
\end{equation}
It means that
$$
\frac{\mathrm{d}\Psi_n}{\Psi_n}=\frac{\mathrm{d}\tilde \Psi_n}{\tilde\Psi_n}\quad
\Rightarrow\quad \ln \Psi_n=\ln\tilde \Psi_n+\mbox{const.}
$$
Hence,
\begin{equation}
	\label{ }
	\tilde \Psi_n(x)=\tilde C\cdot\Psi_n(x).
\end{equation}
We see that two solutions for the bound state with the same energy for the same Hamiltonian differ by a constant. Thus, the bound state has got the only independent wave function: the eigen energy of bound state is non-degenerated. 

For the hermitian Hamiltonian its potential $V$ is real. Therefore, real and imaginary parts of eigen wave functions of Hamiltonian for the bound state are both the solutions of Schr\"odinger equation with the same energy. We have just proven
$$
\mathfrak{Re}\Psi_n=C\cdot\mathfrak{Im}\Psi_n,
$$
hence, one can pose the wave function of bound state to be real in the one-dimensional quantum mechanics. 

\section{Nodes of excited levels \label{sec:AppB}}

\begin{figure}[h!]
	\centering
	\includegraphics[width=0.3\textwidth]{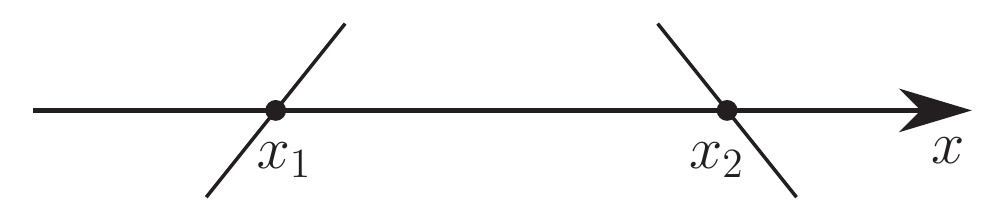}
	\caption{Consecutive nodes and their derivatives.} 
	\label{fig:con_nodes}
\end{figure}

Consider an excited level $\Psi_{k'}$ with number $k'$ greater than number $k$ of level $\Psi_{k}$. Hence,
$$
\left\{\begin{array}{rl}
	\hat H\Psi_{k} &\hskip-3pt= E_{k}\Psi_{k}, \\[1mm]
	\hat H\Psi_{k'} &\hskip-3pt= E_{k'}\Psi_{k'},\\[1mm]
	E_{k'}&\hskip-3pt>E_{k}.
\end{array}\right.
$$
Consider any pair of two consecutive nodes of $\Psi_{k}$ wave function $x_1$ and $x_2$ (if $k=0$ then $x_1$ and $x_2$ are boarder points),
$$
\left\{\begin{array}{rl}
	\Psi_{k}(x_1) &\hskip-3pt= \Psi_{k}(x_2) = 0,\\[1mm]
	\Psi_{k}(x) &\hskip-3pt> 0, \quad x \in (x_1, x_2).
\end{array}\right.
$$
It leads to the following conclusions about the derivatives at points $x_1$ and $x_2$:
$$
\left\{\begin{array}{rl}
	\Psi_{k}'(x_1) &\hskip-3pt> 0,\\[1mm]
	\Psi_{k}'(x_2) &\hskip-3pt< 0,
\end{array}\right.
$$
as it is illustrated on FIG.~\ref{fig:con_nodes}. 

According to proof by contradiction suppose that $\Psi_{k'}(x)>0$ for all $x \in (x_1, x_2)$. This leads to the following statement about the Wronskian:
\begin{equation}\label{minus}
	\int\limits_{x_1}^{x_2}W'_{\Psi_{k}\Psi_{k'}} \, dx= \dfrac{2m}{\hbar^2}(E_{k} - E_{k'}) \int\limits_{x_1}^{x_2}\Psi_{k}\Psi_{k'}\, dx < 0,
\end{equation}
because $E_{k} < E_{k'}$ and both $\Psi_{k}$ and $\Psi_{k'}$ are positive, hence, their integral is positive too.

On the other hand, 
\begin{multline}\label{plus}
	\int\limits_{x_1}^{x_2}W'_{\Psi_{k}\Psi_{k'}}\, dx  =
	W_{\Psi_{k}\Psi_{k'}}\Big|^{x_2}_{x_1}=\\[-2mm] {} = \Psi_{k}'(x_1)\Psi_{k'}(x_1) - \Psi_{k}'(x_2)\Psi_{k'}(x_2) \geqslant 0,
\end{multline}
for continuous wave functions. So we get the contradiction between equations (\ref{minus}) and (\ref{plus}), that means our assumption has been false and there is at least one node of $\Psi_{k'}$ between every two neighboring nodes of $\Psi_{k}$. Therefore, a total number of $\Psi_{k'}$ nodes is greater than a total number of $\Psi_{k}$ nodes, hence, the number of bound state nodes increases with the growth of energy. 
	
	\bibliography{bibl_file}

\end{document}